\documentclass{aa}  
\usepackage{graphicx}
\usepackage{txfonts}
\usepackage{natbib}

\def\degr{\hbox{$^\circ$}}
\def\arcmin{\hbox{$^\prime$}}
\def\arcsec{\hbox{$^{\prime\prime}$}}

\renewcommand{\j}[2]{\mbox{J$= #1\rightarrow #2$}}
\newcommand{\kms}{\mbox{km s$^{-1}$}}
\newcommand{\thirteenco}{\mbox{\rm $^{13}$CO}}
\newcommand{\twelveco}{\mbox{$^{12}$CO}}

\newcommand{\msunyr}{\mbox{M$_\odot$ yr$^{-1}$}}

\newcommand{\Msun}{\mbox{M$_\odot$}}
\def\oversim#1#2{\lower0.5pt\vbox{\baselineskip0pt \lineskip-0.5pt
     \ialign{$\mathsurround0pt #1\hfil##\hfil$\crcr#2\crcr\sim\crcr}}}
\def\lsim{\mathrel{\mathpalette\oversim<}}    % < over \sim
\begin{document}
   \title{The massive expanding molecular torus
   in the planetary nebula NGC~6302}
   \titlerunning{The massive torus of NGC\,6302}
   \author{N. Peretto
          \inst{1}
          \and
          G.~A Fuller \inst{1}
          \and
          A.~A Zijlstra \inst{1}
          \and
          N.~A. Patel \inst{2}
          }

%   \offprints{G. Wuchterl}

   \institute{School of Physics and Astronomy, University of Manchester, Sackville Street, PO Box 88, Manchester M60 1QD, UK
   \and
   Harvard-Smithsonian Center for Astrophysics, 60 Garden Street, MS78, Cambridge, Massachusetts 02138, USA}

\date{}{}             

 %  \date{Received September 15, 1996; accepted March 16, 1997}

% \abstract{}{}{}{}{} 
% 5 {} token are mandatory
 
  \abstract 
   {} 
   {We measure the mass and kinematics of the massive molecular torus in the
     planetary nebula NGC 6302. The nebula is the proto-typical butterfly
     nebula. The origin of the wing-like morphology is disputed: determining
     the mass-loss history of the confining torus is an important step in
     understanding the formation of this structure.  }
  {We performed submillimeter observations with JCMT and the SMA
  interferometer. The continuum emission as well as the J=2--1 and
  3--2 transitions of $^{12}$CO and $^{13}$CO are analysed 
  at arcsecond resolution.  } 
   {The CO emission indicates a mass of the torus
  of $\sim 2$~M$_\odot \pm 1$M$_\odot$.  The $^{12}$CO and $^{13}$CO emission matches the dark
  lane seen in absorption in the H${\alpha}$ image of the object.  The CO
  torus is expanding with a velocity of $\sim8$\,km\,s$^{-1}$, centred at
  $V_{\rm lsr}=-31.5\rm \,km\,s^{-1}$.  The size and expansion velocity of the
  torus indicates that the torus was ejected from $\sim7500$~yr to
  $2900$\,yr ago, with a mass-loss rate  of $5\times10^{-4}\,\rm
  M_\odot\,yr^{-1}$.  We also see a
  ballistic component in the CO images with a velocity gradient of
  140\,km\,s$^{-1}$\,pc$^{-1}$. }  
  {The derived mass-loss history of the torus favours binary
  interaction as the cause of the ejection of the torus. We predict
the existence of
a companion with an orbital period $P\lsim 1\,$month. }


   \maketitle
%
%________________________________________________________________

\section{Introduction}

%The evolution of solar-like stars is terminated by a catastrophic mass
%loss event on the Asymptotic Giant Branch (AGB). During this so-called
%`superwind', between 50\%\ and 80\%\ of the stellar mass is lost, at
%mass-loss rates up to $10^{-4}\,\rm M_\odot\,yr^{-1}$.  The hot
%stellar remnant eventually ionizes its ejected shell: the ionized gas
%is visible as a planetary nebula (PN). 

Asymptotic Giant Branch (AGB) stars lose between 50\%\ and
80\%\ of their stellar mass at mass-loss rates up to $10^{-4}\,\rm
M_\odot\,yr^{-1}$. About 80\% of planetary nebulae (NP), i.e. the
visible part of the ionized gas ejected by AGB stars, show elliptical
and/or bipolar morphologies (e.g. Manchado 1997).
% It is believed that the kinematics of the central star or central binary,
% the outflow, the equatorial enhancement, and the final morphology of the PN
% are closely linked (e.g. Soker 1998; Siess \&\ Livio 1999).
Bipolar PN are believed to form when a fast wind from the now post-AGB star
ploughs into the earlier slow AGB wind, amplifying an initial equatorial
density enhancement (Kwok et al. 1978, Balick et al. 1987; Frank \& Mellema
1994, Zijlstra et al. 2001). Some 15\%\ of PNe show strong morphological
evidence for such an equatorial enhancement of cold gas around the central
star (Corradi \&\ Schwarz 1995).  Detailed studies of these equatorial
structures can then give direct insights on the formation of bipolar PNe.
Related to bipolar nebulae, butterfly PNe do not show well collimated
outflows, but have thicker expanding equatorial torii. The complex morphology
of their outflows has been attributed to an interaction between the fast wind
and a warped disk (Icke 2003). 
% A possiblity (**for what?**) is that the progenitors of bipolar nebulae pass
% through a common envelope stage while butterfly nebulae avoid this (Zijlstra
% 2006).}  The torii observed in the butterfly nebulae are made of relatively
% cold, molecular material, and their detailed study requires then the use of
% millimeter/submillimeter interferometers. This paper explores such an
% interferometric study .

NGC~6302 is the proto-typical `butterfly' nebula. It is located at a
  distance $d\sim 1$~kpc (Meaburn et al., 2005). A broad central absorption
  lane at its centre is suggestive of a massive circumstellar torus.  The
  nebula is driven by one of the highest temperature PN central stars known
  (Casassus et al. 2000), although due to the high extinction of the
  absorption lane, it has not been directly detected.  Previous CO
  observations suggest a mass for the torus of $\sim 0.1$~M$_{\odot}$ (Gomez
  et al. 1989; Huggins \& Healy 1989).
  % Weak CO \j{2}{1} was detected towards this absorption lane by Huggins \&
  % Healy (1989), along with \j{1}{0} emission detected by Gomez et al.
  % (1989).  These observations suggest a total mass for the torus of $\sim
  % 0.1$~M$_{\odot}$.
  However, recently Matsuura et al. (2005a) measured a total gas mass of
  $\sim3$~M$_{\odot}$ from observations of the submillimetre dust continuum
  emission.
% measured a flux of $\sim39$~Jy at 450~$\mu$m 
%, even after correcting for free-free
%  emission, 
  % dust mass of $\sim0.03$~M$_{\odot}$.  Assuming a typical gas to dust mass
  % ratio of 100:1, this implies
  This extremely high mass, especially when compared to the
  $\sim0.7$~M$_\odot$ estimated for the central star (Casassus et al. 2000),
  prompted us to reexamine the CO emission from this source and study of the
  kinematics of the circumstellar gas.

 %__________________________________________________________________

\section{Observations}

\subsection{JCMT observations}

NGC\,6302 was observed with the James Clark Maxwell Telescope
(JCMT)\footnote{The James Clerk Maxwell Telescope is operated by The Joint
  Astronomy Centre on behalf of the Particle Physics and Astronomy Research
  Council of the United Kingdom, the Netherlands Organisation for Scientific
  Research, and the National Research Council of Canada.}  in Hawaii as
observing program S04AU08 on 26 May, 18 and 19 June 2004 using the facility
230~GHz and 345~GHz receivers. We observed \twelveco\ \j{2}{1} and \j{3}{2},
as well as the corresponding transitions of \thirteenco. The data were taken
with a velocity resolution of $\sim 0.1$~km\,s$^{-1}$ for all the transitions,
but was rebinned to 0.4~km\,s$^{-1}$ for analysis.

We raster-mapped the region around NGC~6302 in the \j{2}{1} and
\j{3}{2} transitions of \twelveco. Extended emission components were
seen at some velocities, but the main emission component corresponding
to NGC\,6302 was found to be unresolved at 230\,GHz and only slightly
resolved at 345\,GHz.  We subsequently took \thirteenco\ spectra
pointing towards the $^{12}$CO \j{3}{2} emission peak position, taken
to be $\alpha=17^{\rm h}13^{\rm m}43.9^{\rm s}$,
$\delta=-37^\circ06^{\prime}11^{\prime\prime}$ (J2000).

All observations were made by position-switching to an off position
(0\arcsec,+300\arcsec) from this central position. This off position was
carefully checked from emission which may have contaminated the source
observations, but no emission with a peak temperature of greater than 0.2\,K
(with a line width of 1 km/s or more) was found at the frequency of either the
\j{3}{2}\ or \j{2}{1}\ transitions.  The telescope pointing was regularly
checked during the observations and was found to be constant to within
$\pm2$\arcsec.  A linear baseline has been removed from each of the spectra.
The main beam efficiency of the JCMT is 0.69 and 0.63 at the frequency of the
\j{2}{1} and \j{3}{2} respectively.  The observational parameters are
summarised in Table \ref{obs}.

\begin{table}
  \caption{Observed transitions}             % title of Table
\label{obs}      % is used to refer this table in the text
\centering                          % used for centering table
\begin{tabular}{c c c c c}        % centered columns (4 columns)
\hline                 % inserts double horizontal lines
Transition & Frequency & Telescope & Resolution  & Beam\\ 
           & (GHz)     &           & (km\,s$^{-1}$) & (arcsec) \\%table heading 
\hline                        % inserts single horizontal line
 \twelveco\ \j{2}{1}   & 220.398 & JCMT & 0.4 & 20 \\ 
                       &         & SMA  & 0.3 & $6.2\times 3.2$ \\  
 \thirteenco\ \j{2}{1} & 230.538 & JCMT & 0.4 & 20 \\
                       &         & SMA  & 4.4  & $6.2\times 3.2$ \\  
 \twelveco\ \j{3}{2}   & 345.796 & JCMT & 0.5 & 13 \\   
 \thirteenco\ \j{3}{2} & 330.587 & JCMT & 0.5 & 13 \\
\hline                                   %inserts single line
\end{tabular}
\end{table}

\subsection{SMA observations}

NGC~6302 was observed at a wavelength of 1.3\,mm with the
SMA\footnote{The Submillimeter Array is a joint project between the
Smithsonian Astrophysical Observatory and the Academia Sinica
Institute of Astronomy and Astrophysics, and is funded by the
Smithsonian Institution and the Academia Sinica.}  on 26 June 2005. We
used all 8 antennas in the compact configuration, with a maximum of
baseline of 70\,m. The six shortest baselines range in length from
16.5 to 32 m. The phase center was RA(2000)=17h$^h$13$^m$44.2$^s$
Dec(2000)=-37\degr06\arcmin15.9\arcsec.  The central frequency of the
upper side band was set to 230.538~GHz with a total bandwidth of $\sim
2$~GHz (i.e. covering the frequency range $\sim 229.5$ to $\sim
231.5$~GHz). The lower side band was centered at $\sim 220.5$~GHz
(i.e.  covering $\sim 219.5$ to $\sim 221.5$~GHz). The correlator was
configured to provide a velocity resolution of $\sim
0.26$~km\,s$^{-1}$ for the CO \j{2}{1}\ line and $\sim
4.4$~km\,s$^{-1}$ for the $^{13}$CO \j{2}{1} line.  Weather conditions
were good for 230 GHz observations with a relative humidity of 20\%,
atmospheric opacity at zenith $\tau_{225} = 0.2$ and $T_{sys,
DSB}\approx 250$ K.

The phase and amplitude were calibrated by observing SgrA$^{*}$ for 5 minutes
around every 20 minutes of integration on NGC 6302.  The angular separation
between SgrA$^{*}$ and NGC 6302 is $\sim 11\degr$.  The passband calibration
was done using Callisto, 3C279 and 3C454.3.  The flux calibration was
performed using Callisto assuming a brightness temperature of 120 K at
229.45~GHz. 
The absolute astrometric positional accuracy in the observations
is found to be better than 0.3\arcsec based on observations of the quasars
1911$-$201 and 1924$-$292.

The visibility data were reduced using the SMA version of the Miriad software
package.  Applying natural weighting, the final maps have a synthesized beam
of 6.2\arcsec$\times$3.2\arcsec with P.A.=6.9\degr.  The rms noise in the
continuum image obtained by averaging the LSB and USB data is
$\sigma=5.9$~mJy. The rms noise in the CO channel maps, at 1 km\,s$^{-1}$
resolution, is $\sim 125$ mJy.

\begin{figure}[t]

\includegraphics[width=7.5cm,angle=270]{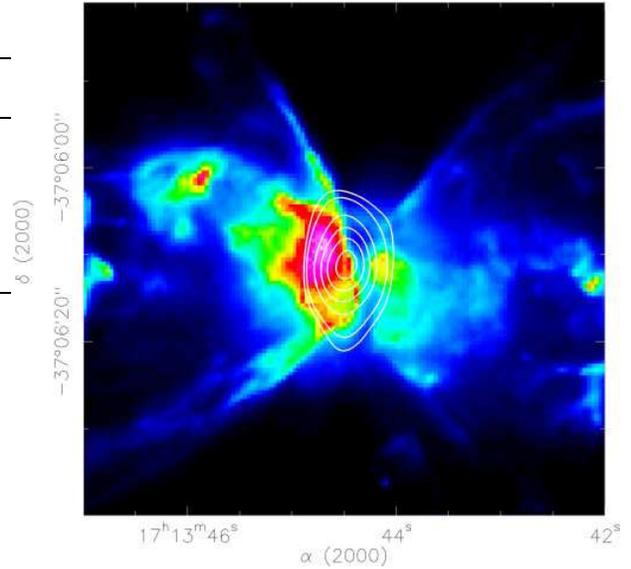}

\caption{
  H${\alpha}$ image from Matsuura et al. (2005a) (colour scale) overlaid with
  our SMA 1.3~mm continuum map (white contours).The contours are 5$\%$ and then from 10
  to 90$\%$ in steps of 20$\%$ of the 1.3~mm continuum emission peak. The emission peak is 0.62 Jy/beam.
\label{cont_map}}
\end{figure}

\section{Continuum emission}

The SMA 1.3~mm continuum map is shown in
Fig.~\ref{cont_map}. To avoid molecular line contamination of
our continuum data, the CO emission, on which the
the correlator bandwidth was centred, has been removed. This continuum map
reveals a single source elongated in the North-South direction. The
angular resolution provided by the SMA spatially resolves the
structure. The position, size and flux are given in Table~\ref{cont},
together with the integrated flux measured inside the 5$\sigma$
contour ($\sim30$~mJy/beam).  

The 1.3~mm continuum flux contains contributions from dust emission
and thermal free--free emission.  The free-free contribution is
$\sim50$\% of the total flux (see Fig.~13 of Matsuura et al. 2005a),
leaving $0.65$~Jy from the dust continuum emission. Assuming
a mean dust temperature ranging from 30~K to 120~K (Kemper et
al. 2002), a dust to H$_2$ mass ratio of 1$\%$, a typical dust opacity
for protostellar disks of 2~cm$^2$\,g$^{-1}$ (Beckwith et al. 1990),
and a distance $d\sim 1$~kpc (Meaburn et al., 2005), we infer a total
(dust plus gas) mass ranging from 0.8~M$_{\odot}$ to
3.9~M$_{\odot}$. However, the temperature range given by Kemper et
al. (2002) corresponds to two different components, a cold one with
T$_k =$ 30 to 60~K, and a hot one with T$_k =$ 100 to 120~K. As
discussed later (see Sect.~5.3), the bulk of the mass comes from the
cold component, meaning that 0.8~M$_{\odot}$ is rather a lower limit
to the mass. Taking 60~K as the average temperature, we find a mass of
1.8~M$_{\odot}$. Uncertainties due to temperature and dust opacity are
at least a factor of 2 in each way.  Comparing this mass estimate to
the one from Matsuura et al. (2005a) obtained from single dish
observations, we conclude that there is almost no missing dust
continuum flux in our SMA data, as confirmed later in Sec.~5.1 through
our molecular line data set. Because the SMA is not sensitive to
extended emission larger than approximately 20\arcsec\ the lack of
missing flux confirms then the compact nature of the NGC~6302 torus.

The peak of the millimetre map is slightly offset, i.e. $\sim
2$\arcsec, from the centre of the dark lane.  Two effects explain this
offset.  First, the contribution from the free-free emission seen at
6~cm (Matsuura et al. 2005a) shifts the 1.3~mm peak toward the location of
the free-free emission peak. Second, the dust absorption is tracing
only the foreground part of the dusty torus, whilst the 1.3~mm
emission traces it entirely. A slight tilt of the torus results in the
1.3~mm emission appearing offset from the absorption.

\begin{table*}[t]
  \caption{SMA 1.3~mm dust continuum and CO 2-1 results. $S_{\rm int}$
is the integrated flux within the 5$\sigma$ contour. The FWHM is the
deconvolved Gaussian size, assuming position angle is zero. } 
\label{cont}      
\centering                       
\begin{tabular}{c c c c c c}        
\hline\hline                 % inserts double horizontal lines
Tracer & Coordinates & S$_{peak}$ & S$_{int}$ & FWHM  \\
&(J2000) & (Jy/beam) & (Jy) & (arcsec) \\%table heading 
\hline                        % inserts single horizontal line
  Cont. 1.3mm & 17:13:44.45~ $-$37:06:11.1  & 0.62 & 1.26 & $6.0\times4.6$\\   
  CO \j{2}{1} & 17:13:44.47~ $-$37:06:08.0  &   &  & $11.3\times4.2$\\   
\hline                                   %inserts single line
\end{tabular}
\end{table*}

\section{Line data}

\subsection{Averaged spectrum}

\begin{figure}[!t!]
\includegraphics[width=8cm,angle=0]{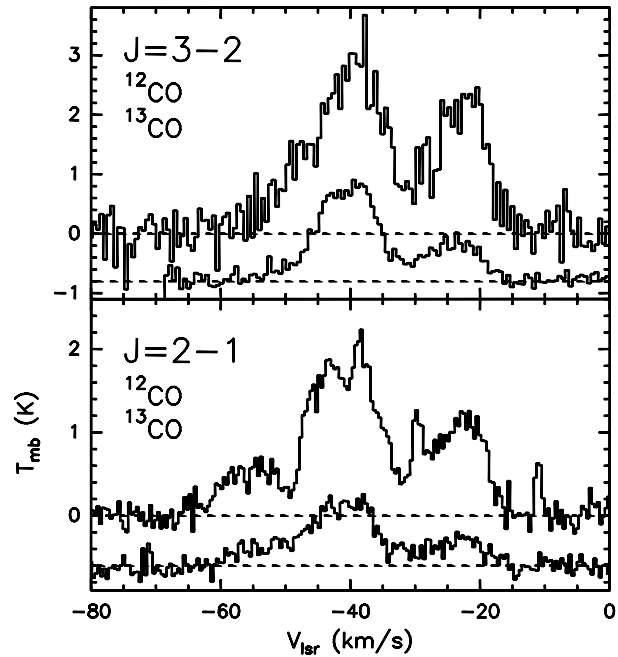}
\caption{$^{12}$CO and $^{13}$CO spectra obtained with the JCMT at the
  position RA(2000)= 17$^h$13$^m$43.9$^s$
  Dec(2000)=-37\degr06\arcmin11\arcsec. Note the presence of three velocity
  components in each line. 
  \label{spec_jcmt}}
\end{figure}

The JCMT data show a compact CO structure approximately coincident with the
continuum emission, which traces the molecular content. The CO spectra for all
four transitions, at the central position, are shown in Fig.~\ref{spec_jcmt}.
The spectra show several broad components, with emission ranging between $-15$
and $-65\,\rm km \,s^{-1}$, using local standard of rest (LSR) velocities.  In
addition, several narrow components are seen in \twelveco\ only, and are
better seen in \j{2}{1} than in \j{3}{2}. These components are at $-38$, $-30$
and $-10\,\rm km \,s^{-1}$. The relative strengths of these features suggest
they trace gas which has a high \twelveco/\thirteenco\ ratio, and may be
spatially extended.  

\begin{figure}[!t!]
\includegraphics[width=4.5cm,angle=270]{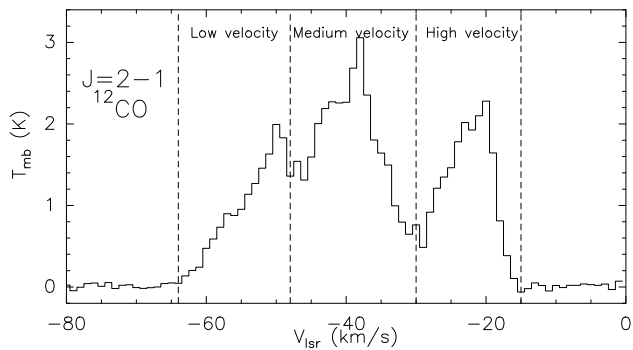}
\caption{Averaged $^{12}$CO \j{2}{1}\ spectrum  obtained with the SMA,
 integrated over the full spatial extent of the source. 
The dashed lines mark the velocity range of each of the velocity components.
The stellar velocity, estimated at $-31.5 \,\rm km\,s^{-1}$ (cf Sec.~6.2), is approximately
the boundary between the medium and high velocity components.
\label{spec_sma}}
\end{figure}

The SMA has both higher resolution, and as an interferometer it is not
sensitive to very extended emission. This allows us to identify interstellar
emission components, which will be absent from the SMA maps. The averaged
\twelveco\ \j{2}{1} obtained with the SMA is shown in Fig.~\ref{spec_sma}
\footnote{We have applied a conversation factor of 0.8~K/Jy, which takes into
  account the frequency and the size of the beam, to convert the SMA line
  observations to a main beam temperature scale.}. Two of the narrow
components are absent, but the one at $-38\,\rm km \,s^{-1}$ is seen,
suggesting this gas is associated with the molecular torus. Its weakness at
\j{3}{2} in the JCMT data can be attributed to an off-centre position combined
with the smaller beam at this frequency.
The narrow components at $-30$ and $-10\,\rm km \,s^{-1}$, non observed on our SMA data,  may be unrelated interstellar gas, or possibly associated with the extended (8 arcmin) emission of the planetary nebula.

Based especially on the averaged spectrum, we define three broad velocity
components, at $\sim-25$ \kms, $\sim-40$ \kms\ and a feature/wing at $\sim-50$
to $-60$ \kms. These features are present in all four spectra, but are best
defined at \twelveco\ \j{2}{1}. The same three components are also seen in
the CO \j{1}{0} of Gomez et al. (1989). The low velocity feature is much stronger in
the SMA integrated spectrum than in the \j{2}{1} JCMT spectrum, indicating it
is located off-centre. It is absent from the \j{3}{2} JCMT spectrum.

\subsection{SMA $^{12}$CO \j{2}{1} and $^{13}$CO \j{2}{1} maps}

The SMA channel maps allow us to define the precise velocity ranges for each
of the three broad components.  These range from $-64$~\kms\ to $-48$~\kms\
for the low velocity component, from $-48$~\kms\ to $-30$~\kms\ for the medium
velocity component, and from $-30$~\kms\ to $-15$~\kms\ for the high velocity
component. The relative intensity of these components varies from position to
position, resulting from a complex velocity structure in the source.

\begin{figure}[!t!]
\includegraphics[width=8cm,angle=270]{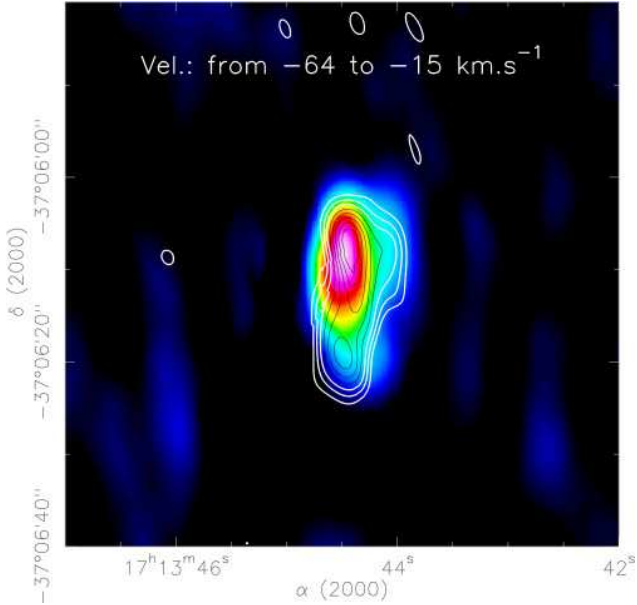}
\caption{ Superposition of the $^{12}$CO \j{2}{1} integrated intensity map  (colour scale) with the $^{13}$CO \j{2}{1} integrated intensity map (contours), both integrated over the full velocity range, i.e. from -64 to -15~km s$^{-1}$. The peak value of the $^{12}$CO \j{2}{1} emission is 190 K km s$^{-1}$, while for the $^{13}$CO \j{2}{1} emission the contours go from 20 to 90~$\%$ in steps of 10$\%$, with a peak at 78 K km s$^{-1}$. 
\label{velcompco21}}
\end{figure}

 The $^{13}$CO \j{2}{1}\  transition is optically thinner, and then is a better tracer of the mass distribution and kinematics than the $^{12}$CO \j{2}{1}. In the following we then discuss further the $^{13}$CO \j{2}{1}\ emission. Figure~\ref{velcompco21} displays the full integrated $^{12}$CO \j{2}{1}\  (colour scale) and $^{13}$CO \j{2}{1}\ (contours) emission
observed with the SMA. We can see that the $^{13}$CO \j{2}{1}\ emission is more structured and elongated  than the $^{12}$CO \j{2}{1}\ one. 
Comparing with the continuum emission, we see that the line emission is more extended, and
peaks three arcseconds to the North of the continuum peak (see Table
\ref{cont}). The lower contour levels also show extensions to the
South and East which are not as strong in the continuum.  Part of the
difference may be caused by the free-free contribution to the
continuum.

\begin{figure*}[!ht!]
\hspace{1cm}
\includegraphics[width=12cm,angle=270]{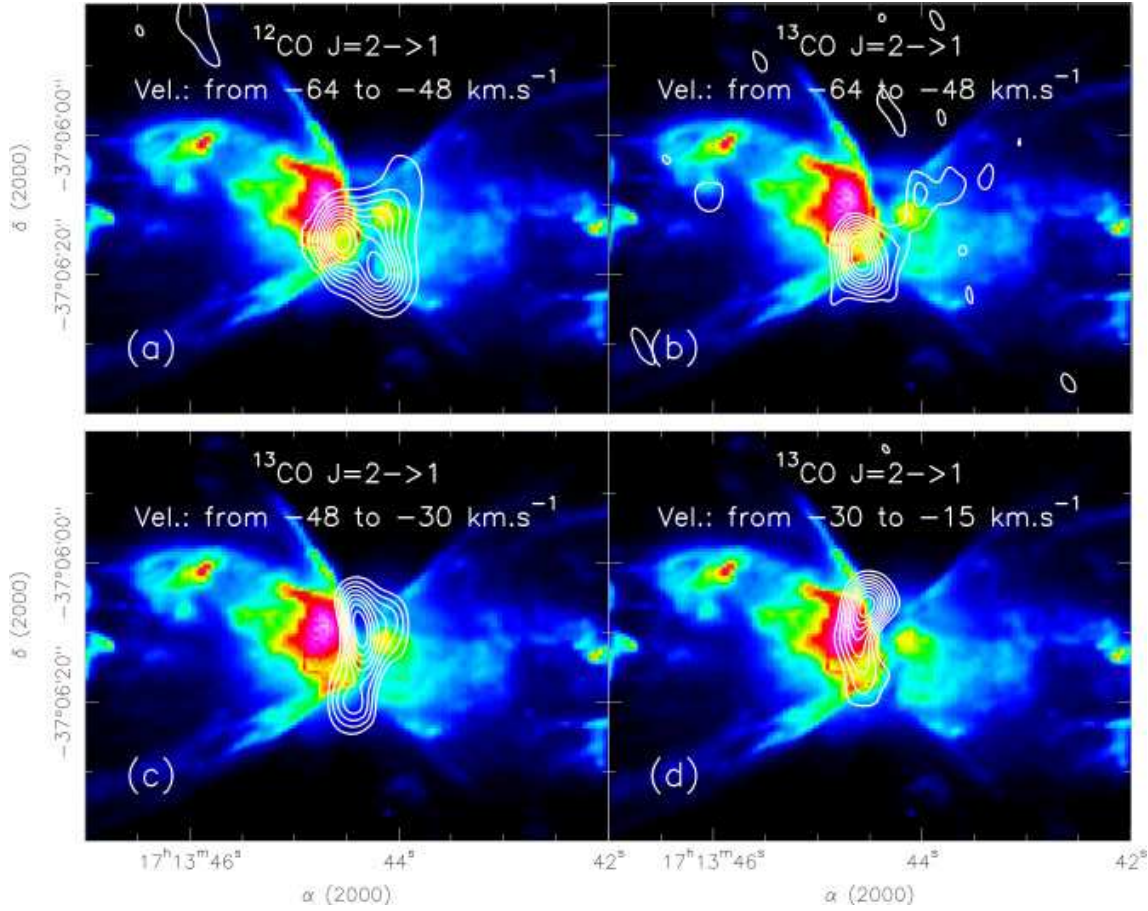}
\caption{Superposition of the H$\alpha$ image (colour scale) with CO integrated intensity maps (white contours) obtained with the SMA.
 In each panel the contours go from 20 to 90~$\%$ of the
  emission peak in steps of 10$\%$. \textbf{a)} $^{12}$CO \j{2}{1} integrated intensity map from -64 to --48~km\,s$^{-1}$ with a peak value of 47 K km\,s$^{-1}$; \textbf{b)} Same as (a) for the $^{13}$CO \j{2}{1}\ transition. The emission peak is 26~K\,km\,s$^{-1}$; \textbf{c)} $^{13}$CO \j{2}{1}\ integrated intensity map
  from $-47$ to $-30$~km\,s$^{-1}$, the emission peak is 68~K\,km\,s$^{-1}$;
  \textbf{d)} $^{13}$CO \j{2}{1}\ integrated intensity map from $-29$ to $-15$~km\,s$^{-1}$, the emission peak is
  43~K\,km\,s$^{-1}$.
\label{velcomp13co21}}
\end{figure*}

 Fig.  \ref{velcomp13co21} shows the low velocity interval of the \twelveco\ \j{2}{1}\ (a) and \thirteenco\ \j{2}{1}\  (b), as well as the medium and high intervals of the \thirteenco\ \j{2}{1}\ line (b and d). The low velocity component (Fig.~\ref{velcomp13co21}a and b) differs
significantly from one line to another. The $^{13}$CO shows only one clear
component, which is offset to the South-East compared to the brighter
(and closest) \twelveco\ component. The missing emission peak in the $^{13}$CO is likely
too weak to have been detected in these observations which have a one
$\sigma$ noise level of $\simeq2$~K~\kms. Its weakness indicates a
lower opacity for this clump, which is not apparent in $^{12}$CO
\j{2}{1} where both clumps are optically thick.  The cause for the
position shift is less clear, but it may indicate a difference between
the location of the temperature peak, traced by the optically thick
\twelveco\ peak, and the column density peak, traced by the optically
thin \thirteenco~peak.

The medium velocity component (Fig.~\ref{velcomp13co21}c)
matches the dark lane seen in the H${\alpha}$ map better than the
$^{12}$CO, and leaves little doubt about its association with the dark
lane. We note also a slightly curved shape of this component. The high
velocity component (Fig.~\ref{velcomp13co21}d) is more extended than
its $^{12}$CO counterpart, and is curved in the opposite direction to
the medium velocity component.  The lack of evidence for dust
extinction associated with this strong CO component suggests it is
located behind the H${\alpha}$ emitting material.  The medium
and high velocity peaks are situated on either side of the continuum
peak, but shifted North by a few arcseconds. The location and
velocity of these two components suggests they belong to an expanding,
toroidal structure, but the northward shift suggests the torus may be
weaker or incomplete towards the South.

In the next section we will discuss the mass of this torus based on CO observations

\section{Mass determination}
\label{sec:mass}

\begin{table*}
\caption{JCMT $^{12}$CO and $^{13}$CO integrated intensities for the three 
velocity components in $T_{\rm mb}$ scale. The values in parenthesis are the one $\sigma$ standard deviation uncertainties.}        
\label{co_prop}     
\centering          
\begin{tabular}{ c c c c c c }   
\hline\hline               
Source Name & Velcocity range  & $I_{\rm int}^{^{12}\rm CO(3-2)}$ & 
$I_{\rm int}^{^{12}\rm CO(2-1)}$ & 
$I_{\rm int}^{ ^{13}\rm CO(3-2)}$ &  
$I_{\rm int}^{ ^{13}\rm CO(2-1)}$\\
& (km\,s$^{-1}$) & (K\,km\,s$^{-1}$) & (K\,km\,s$^{-1}$) & (K\,km\,s$^{-1}$) &
(K\,km\,s$^{-1}$)\\
\hline   
low vel. comp. &   -64 to -48  & 20.0 (2.4)  &12.5 (1.4) & 1.8 (0.3)  & 2.6 (0.2)\\ 
medium vel. comp. &  -48 to -30 &  56.1 (2.6) & 27.3 (1.5) & 17.4 (0.3)& 8.7 (0.2)\\ 
high vel. comp. &  -30 to -15 & 40.7 (2.4)& 15.4 (1.4) & 6.2 (0.3) &  3.0 (0.2)\\
 
      % inserting body of the table
\hline    
\end{tabular}
\end{table*}

In order to determine the total molecular mass of the torus we
  estimate the excitation temperature and opacity of the CO lines
  using two methods. These both rely on a good knowledge of the
  observed $^{12}$CO and $^{13}$CO line ratios.

Since our JCMT $^{12}$CO observations only barely resolve the torus,
even in the \j{3}{2} transition, we assume that the central JCMT
spectra contains all the emission from the torus and so we can use
these spectra alone to determine the total molecular mass.  However,
note that this is not true for the low velocity component which is not
part of the torus, as this component peaks at 7\arcsec\ South-East.

Table~\ref{co_prop} gives the JCMT main beam integrated intensities of the
four transitions, $^{12}$CO \j{3}{2}, $^{12}$CO \j{2}{1}, $^{13}$CO
\j{3}{2}, and $^{13}$CO \j{2}{1}, integrated over the three velocity
components identified in our SMA data.
As noted above the CO \j{3}{2} intensity is likely underestimated for
the low velocity component.
The JCMT beam filling factors can be calculated quite accurately for
each velocity component using the source size measured in our SMA
observations. These beam filling factors are listed in
Table~\ref{fil_fac} assuming that the size of the source is the same
for both CO species and both transitions. The filling factor
corrected line ratios are given in Table~\ref{ratio}. Also, the comparison of the JCMT and SMA $^{12}$CO \j{2}{1}\
integrated flux shows that the SMA observations recover $\sim 80\%$ of
the JCMT flux, demonstrating once again the compact nature of NGC~6302.  

\begin{table}
\caption{Beam filling factors (BFF)}             % title of Table
\label{fil_fac}      % is used to refer this table in the text
\centering                          % used for centering table
\begin{tabular}{c c c   }        % centered columns (4 columns)
\hline\hline                 % inserts double horizontal lines
Source Name & BFF CO \j{2}{1} & BFF CO \j{3}{2}  \\
\hline                        % inserts single horizontal line
low vel. comp. &  0.62 &  0.93 \\ 
medium vel. comp.  & 0.43 &  0.64 \\ 
high vel. comp. & 0.31 & 0.57 \\
 
      % inserting body of the table
\hline                                   %inserts single line
\end{tabular}
\end{table}

\begin{table*}
\caption{Integrated line ratios. The values in parenthesis are the one $\sigma$ standard deviation uncertainties.}             % title of Table
\label{ratio}      % is used to refer this table in the text
\centering                          % used for centering table
\begin{tabular}{c c c c c  }        % centered columns (4 columns)
\hline\hline                 % inserts double horizontal lines
Source Name & 
$I_{\rm int}^{^{12}\rm CO(3-2)}$/$I_{\rm int}^{^{12}\rm CO(2-1)}$&
 $I_{\rm int}^{^{13}\rm CO(3-2)}$/$I_{\rm int}^{^{13}\rm CO(2-1)}$& 
$I_{\rm int}^{^{12}\rm CO(3-2)}$/$I_{\rm int}^{^{13}\rm CO(3-2)}$&
$I_{\rm int}^{ ^{12}\rm CO(2-1)}$/$I_{\rm int}^{ ^{13}\rm CO(2-1)}$\\
\hline                        % inserts single horizontal line
low vel. comp. & 1.07 (0.25) &  0.46 (0.11)&  11.11 (3.18) & 4.81 (0.91)\\ 
medium vel. comp. & 1.38 (0.14)&  1.34 (0.05) &  3.22 (0.21) &  3.14 (0.24)\\ 
 high vel. comp.& 1.35 (0.22)&  1.12 (0.13) &  6.56 (0.70) & 5.13 (0.81) \\
 
      % inserting body of the table
\hline                                   %inserts single line
\end{tabular}
\end{table*}

\subsection{Analytical estimates of the kinetic temperature and opacity}

The ratio of two different transitions of the same molecule
  constrains the excitation temperature, as given by Eq.(3) of
  Levreault (1988), assuming that the beam filling factors and the
  opacities of the transitions are known.  For simplicity, we assume
  that the excitation temperature is uniform along the line-of-sight
  and that the transitions are thermalized, which allows us to infer
  the opacity of one transition given the opacity of the other one.

For the NGC~6302 case, we know that the $^{13}$CO lines are not
  optically thick (i.e. $\tau < 1$).  So assuming an opacity of 0.5
  for the $^{13}$CO \j{3}{2} line we find a kinetic temperature of
  $T_{\rm k} = 25\pm5$\,K for the high velocity component and $T_{\rm
    k} = 35\pm5$\,K for the medium velocity component.  For the low
  velocity component we find $T_{\rm k} \le 10$\,K, but because of the
  missed flux in the $^{13}$CO \j{3}{2}\ spectrum
  (Sec.~\ref{sec:mass}) we consider this value underestimated by a factor
  of $\sim2$ and believe that kinetic temperature of $T_{\rm k} \simeq
  20$\,K is more realistic for this component.

With this range of kinetic temperatures inferred from the $^{13}$CO ratio, it
is then possible to infer the opacity of the $^{12}$CO lines.  We find a
maximum opacity for the $^{12}$CO \j{3}{2} transition of unity for every
velocity component, taking into account the
uncertainties on the observed ratio.\\

The opacity and kinetic temperature can also be inferred from the line
ratio of two isotopic substituted molecules for the same transition.
This method does not require the knowledge of the beam filling factor,
if the filling factor is the same for both species, but it does
require knowledge of the relative abundance of the species used.
Assuming that the beam filling factor is the same for both
transitions, that the excitation temperature is the same for both
transitions, and also that the linewidth is the same in both
transitions, the ratio of the integrated intensities is given by
Eq.(3) of Myers et al. (1983).

Although the terrestrial value of $A_b$, the abundance ratio
$X[^{12}\rm CO]$/$X[^{13}\rm CO]$, is $90$, for PNe $A_b < 30$ (Balser
et al. 2002), and 15 is a lower limit for NGC~6302
(Sec.~\ref{sec:twocomp}).  For $A_b=15$ the observed
$^{12}$CO(2-1)/$^{13}$CO(2-1) ratios lead to a $^{12}$CO \j{2}{1} line
opacity of 3 for the high and low velocity components and 5.5 for the
medium velocity component, with a typical uncertainty of $\pm0.5$. The
corresponding opacities for the $^{12}$CO \j{3}{2} transitions are
then $3\pm0.5$ and $7.5\pm0.5$ for the high and medium velocity
components, respectively. The opacity inferred for the low velocity
component is $\sim 1\pm1.5$, well below the one inferred from the
\j{2}{1} transitions. This is likely due to the issue with the
$^{13}$CO \j{3}{2} already mentioned in the previous section. The corresponding opacities in the
$^{13}$CO transitions are thus 15 times smaller, that is $\sim0.1$ for
the low velocity component, 0.2 for the high velocity component, and
$\sim0.5$ for the medium velocity component, with a typical
uncertainty of $\pm0.05$. For an abundance ratio $A_b=30$ rather than
15, the opacities are approximately doubled.

The $^{12}$CO opacities inferred through this method are 3 to 15 times
higher than the ones inferred using two transistions of the same
species. Such differences in the opacity estimates lead to a similar
dispersion in the mass estimates and indicate the need for a more
complete model.

\begin{figure}
\includegraphics[width=6.5cm,angle=270]{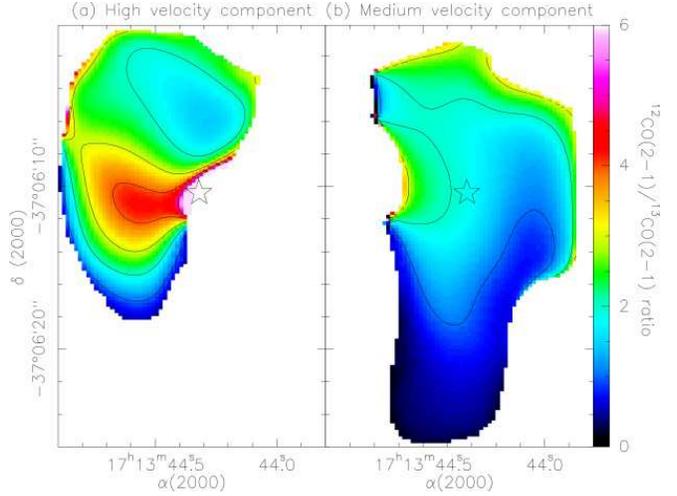}
\caption{$^{12}$CO(2-1)/$^{13}$CO(2-1) ratio from the SMA data. The star 
symbol shows the assumed  position of the central star
(Matsuura et al. 2005a). \textbf{a)} High velocity component. 
\textbf{b)} Medium velocity component.
\label{ratio_map}}
\end{figure}

\subsection{A two phase model: A hot inner edge and a cold outer part}
\label{sec:twocomp}

The most likely explanation for the discrepancy between the two
analytical methods discussed above is the presence of opacity and
temperature gradients within the beam (Levreault 1988).  Given the
presence of a luminous, hot stellar remnant in the centre of NGC~6302,
such gradients would not be unexpected.

Figure~\ref{ratio_map} shows maps of the ratio
$^{12}$CO(2-1)/$^{13}$CO(2-1) for the medium and high velocity
components. For both components this ratio is far from uniform over
the structure. The high velocity component shows higher values of this
ratio than does the medium velocity component; the peak is located in
the central part, close to the proposed location of the central star
(Matsuura et al. 2005a).  Direct or indirect energy input by the
central star appears to be responsible for a temperature gradient
within the molecular structure, making the gas hotter in the inner
part and colder in the outer part.

Figure \ref{ratio_map} suggests an inner, marginally optically thin,
warm component, seen directly through the high velocity component, and
an outer, optically thicker, cold part, obscuring the warm component
associated with the medium velocity component. This is consistent with
the interpretation above that the high velocity component is located
behind the star, while the medium velocity component is in front of
star (Sec.~4.2).

The highest value of the $^{12}$CO(2-1)/$^{13}$CO(2-1) ratio, found
for the low velocity component, gives a lower limit to the abundance
ratio $A_b$ of 15.

To account for the observed non-uniform physical conditions we have
modelled the CO emission assuming it arises from two phases with
different temperatures. Using the 1D radiative transfer code RADEX
(Sch\"oier et al. 2005) we have constructed a grid of models
varying the kinetic temperature and the column density of the CO. The
volume density, the linewidth and the background temperature were
fixed to $1\times10^5$~cm$^{-3}$, 10~\kms\ and 2.73~K, respectively.
We used two different values, 15 and 30, for the abundance ratio
between $^{12}$CO and $^{13}$CO. Four grids of models were
constructed, one for each transition, $^{12}$CO \j{2}{1}, $^{12}$CO
\j{3}{2}, $^{13}$CO \j{2}{1}, and $^{13}$CO \j{3}{2}.  Each individual
model within these grids gives the resulting radiation temperature for
a medium with a different column density and kinetic temperature. The
$^{12}$CO column density was varied from $5\times10^{16}$~cm$^{-2}$ to
$5\times10^{18}$~cm$^{-2}$ in steps of $5\times10^{16}$~cm$^{-2}$, and
the kinetic temperature was varied from 10~K to 300~K in steps of 5~K.
In order to simulate a two-phase medium we mixed each point on each
grid with every other point, adopting a contribution of each phase to
the observed spectra ranging from 10$\%$ to 90$\%$ in steps of 10$\%$.

This procedure generated almost $6\times10^7$ different two component
models.  We then built for each model, the four line ratios, i.e.
$^{12}$CO(3-2)/$^{12}$CO(2-1), $^{13}$CO(3-2)/$^{13}$CO(2-1),
$^{12}$CO(3-2)/$^{13}$CO(3-2), and $^{12}$CO(2-1)/$^{13}$CO(2-1).  A
$\chi^2$ analysis on these grids yields the models which best fit the
observed ratios for each of the three velocity components.

\begin{figure}[!t!]
\includegraphics[width=5cm,angle=270]{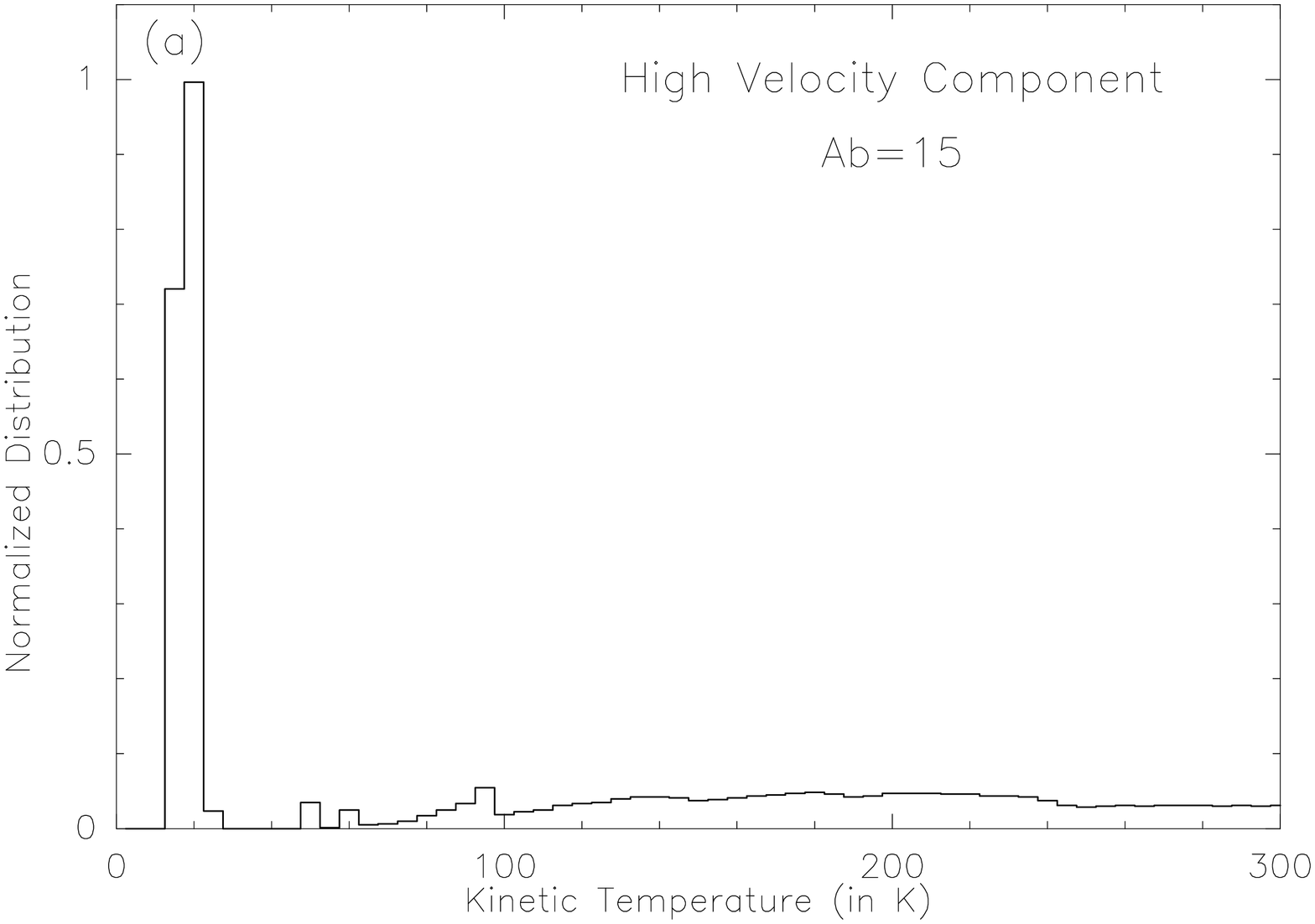}

\vspace{0.3cm}
\includegraphics[width=5cm,angle=270]{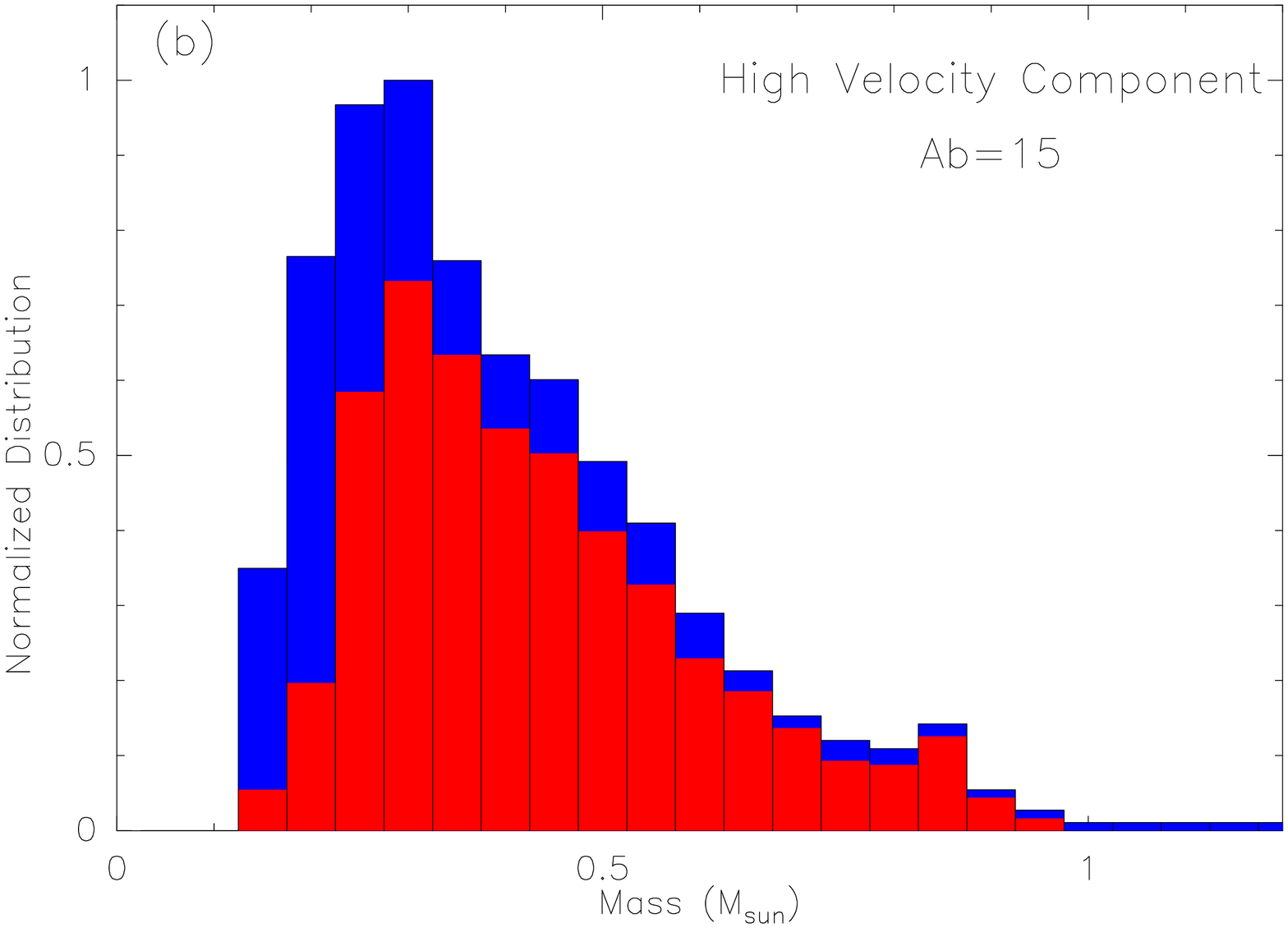}
\caption{Results of radiative transfer modelling of a two phase medium
  for the high velocity component. \textbf{a)} Temperature
  distribution of the models fitting the four observed line ratios
  within the uncertainties. These models use an abundance ratio
  $X$[$^{12}$CO]/$X$[$^{13}$CO] of 15. \textbf{b)} Mass distributions of
  the models fitting the four observed line ratios within the
  uncertainties. The fraction of hot material (i.e. with $T_{\rm k} > 80$~K)
  is represented in red, while the fraction of cold material (i.e.
  with $T_{\rm k} < 80$~K) is represented in blue.
\label{model_high}}
\end{figure}

Figure~\ref{model_high} shows the results of the radiative transfer
modelling for high velocity component. These plots only shows the
model ratios fall within the uncertainties of the observed ratios
(Table~\ref{ratio}), for the preferred abundance ratio of 15. We
can clearly see on Fig~\ref{model_high}a that two different phases are
needed to reproduce the observed ratios, one cold phase, with a
kinetic temperature $\sim20$~K and a warmer phase, with a kinetic
temperature ranging from $\sim$80~K to 300~K, depending on the model.

Figure~\ref{model_high}b shows the mass distribution of these models. For each
mass bin we show in red and blue the relative percentage of hot and cold
material, respectively. Here, hot means  a kinetic temperature of
$T_{\rm k}>80$\,K and cold means $T_{\rm k}<80$\,K. The most likely mass for
the high velocity component is $\sim 0.3$\,M$_{\odot}$ and this mass is
largely dominated by hot molecular gas. This is consistent with the fact that
the high velocity component exhibits the hot inner part of the torus.

\begin{figure}[!t!]
\includegraphics[width=5cm,angle=270]{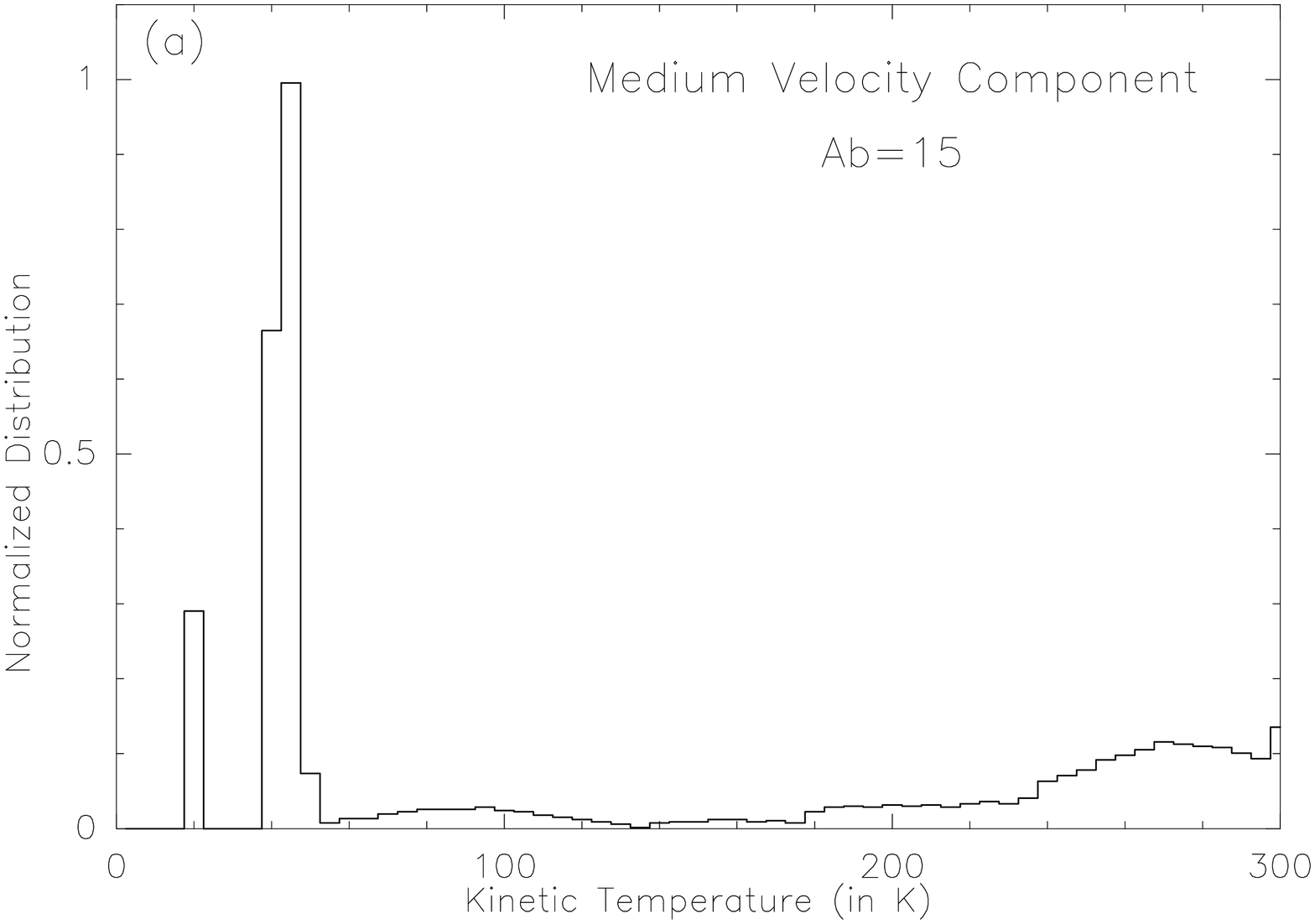}

\vspace{0.3cm}
\includegraphics[width=5cm,angle=270]{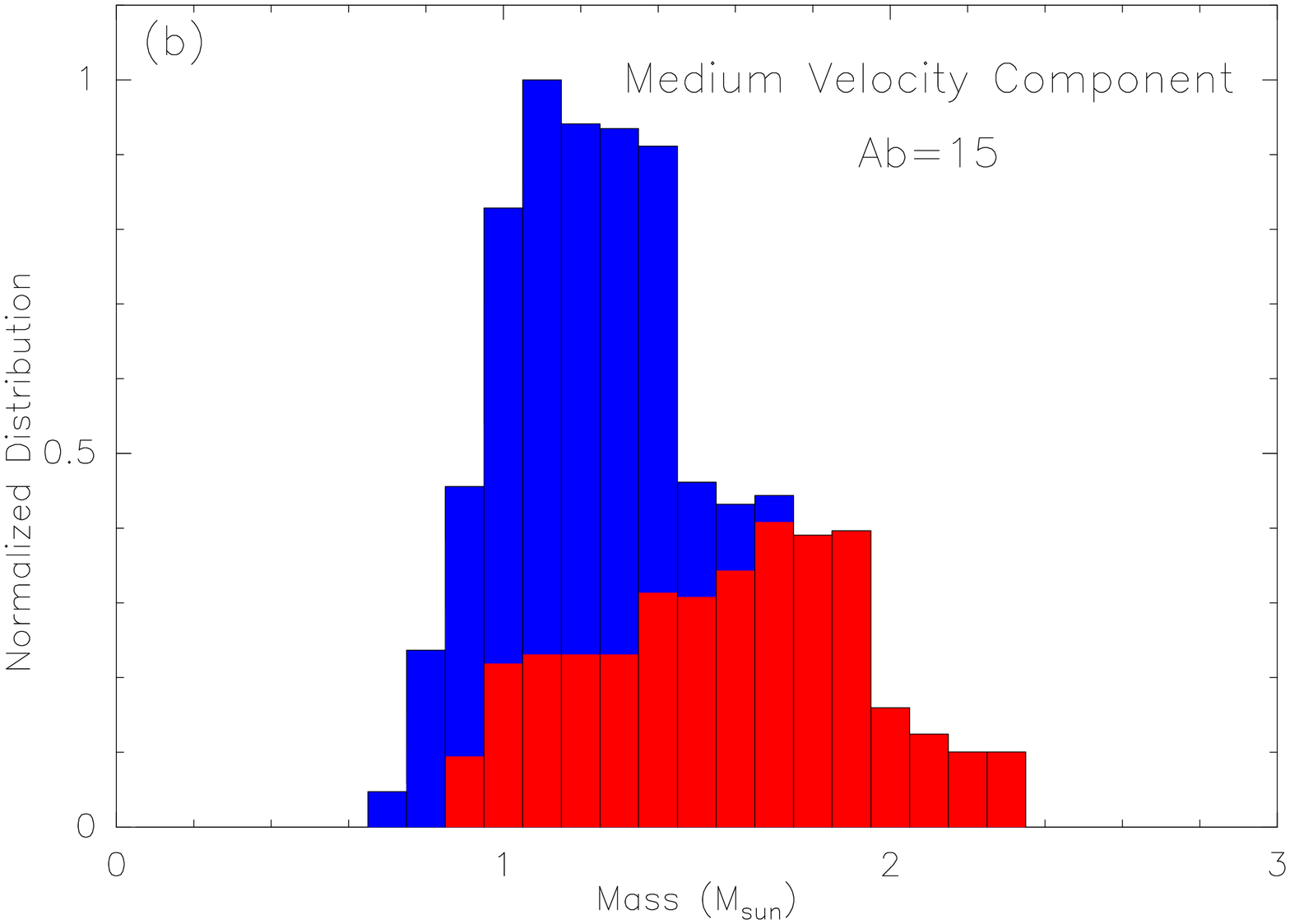}
\caption{Same as for Fig.~\ref{model_high} but for the medium velocity 
component.
\label{model_med}}
\end{figure}

 Figure~\ref{model_med} shows the same plots for the medium velocity component.
We see that the temperature distribution is quite similar to that of the high
velocity component, but the mass distribution is very different. The most
likely mass for the medium velocity component is $\sim 1.1$~M$_{\odot}$ and
this mass is this time largely dominated by cold material. This is consistent
with our picture of NGC~6302 since the medium velocity component 
shows the cold outer part of the torus. Using an
abundance of 30 increases the best fit masses by roughly a factor of 2, but we
did not find any model for the medium velocity component fitting within the
uncertainties of the four line ratios, suggesting 15 is a better choice value
for the abundance ratio.
 
The derived mass of the medium velocity component is almost four times
higher than the mass of the high velocity component.  This can be
explained by the optical depth effects: the bulk of the cold phase of
the high velocity component is hidden by the hot optically thick
material in front of it, and thus, does not appear in our mass
estimate of the high velocity component.  However this cold component
actually dominates the total mass of the toroidal structure. This
discrepancy shows the limits of our simple two phase model.  Although
the total mass of the molecular content resulting from our models is
1.4~M$_{\odot}$, geometrical symmetry suggests that the actual mass is
closer to 2~M$_{\odot}$, with an estimated uncertainty of
  1~M$_{\odot}$ given the mass distribution of the fit models
  (Fig.~\ref{model_high}b and \ref{model_med}b). This value is in
very good agreement with the mass estimate from the dust continuum
emission (see Sect.~3).

We tried to estimate the mass of the low velocity component in
the same way, but excluding the somewhat unreliable $^{13}$CO \j{3}{2}
data for this component did not allow us to identify a clear mass
limit.  For this component we have estimated a lower limit on the mass
by assuming a temperature T$_k = 20$~K, LTE, and using the opacity of
the $^{13}$CO \j{2}{1} line, $\sim0.35$ (obtained in a same way as for
the $^{12}$CO \j{2}{1} opacity; cf Sect.~5.1). Combining equations (9)
and (19) from Goldsmith \& Langer (1999), and using the $^{13}$CO
\j{2}{1} integrated intensity in Table~\ref{co_prop} we find a lower
limit of 0.06~M$_{\odot}$ for the mass of the low velocity component.
All these mass estimates are summarized in Table~\ref{mass}.

\begin{table}
\caption{Column densities and mass. The mass of the low
velocity component is considered a lower limit. The mass of
the high velocity components is only likely underestimated}      
\label{mass}      % is used to refer this table in the text
\centering                          % used for centering table
\begin{tabular}{c c c c    }        % centered columns (4 columns)
\hline\hline                 % inserts double horizontal lines
Component & N$_{^{13}CO}$& N$_{H_2}$& Mass  \\
 & (10$^{16}$~cm$^{-2}$)& (10$^{21}$~cm$^{-2}$)& (M$_{\odot}$)\\
\hline                        % inserts single horizontal line
high velocity &  2.9 &  4.4 & 0.3\\ 
medium velocity  & 10.7  &  16.1 & 1.1\\ 
low velocity  & 0.3 & 0.5  & 0.06\\
 
      % inserting body of the table
\hline                                   %inserts single line
\end{tabular}
\end{table}

\section{Kinematics}

The molecular structure surrounding the central star of NGC~6302 is
elongated in the North-South direction (see Fig.~\ref{velcompco21} and
\ref{velcomp13co21}) and is coincident with the elongated structure
previously seen in absorption on the H${\alpha}$ image (Matsuura et
al. 2005a).  When we overlay the high and medium velocity components
observed in CO with the SMA (see Fig.~\ref{ellipse}a and c), we see
that these two compoments appear to form the farside and nearside
parts of a thick disk-like shaped structure surrounding the central
star. Based on the angular size observed in $^{13}$CO \j{2}{1} (see
Fig.~\ref{ellipse}c) we can estimate the physical diameter and
thickness of this structure.  Considering that NGC~6302 is located at
$d\simeq1$~kpc (Meaburn et al. 2005), we infer a diameter for the
disk-like structure of D$\simeq 0.11$~pc, and a thickness $T\simeq
0.04$~pc. However although the morphology is similar to a disk, the
velocity structure is clearly not.

\subsection{The torus velocity structure}

Considering the medium and high velocity CO components as tracing a disk-like
structure, the velocity channel marking the limit between the two should
correspond to the systemic velocity of NGC~6302. This velocity is found to be
about $V_{\rm lsr} = -30\pm 1$~km\,s$^{-1}$.  This is in remarkable agreement
with the value of $-30.4$~km\,s$^{-1}$ derived by Meaburn et al. (2005) given
the very different methods used.

\begin{figure}
\hspace{1cm}
\includegraphics[width=5.5cm,angle=270]{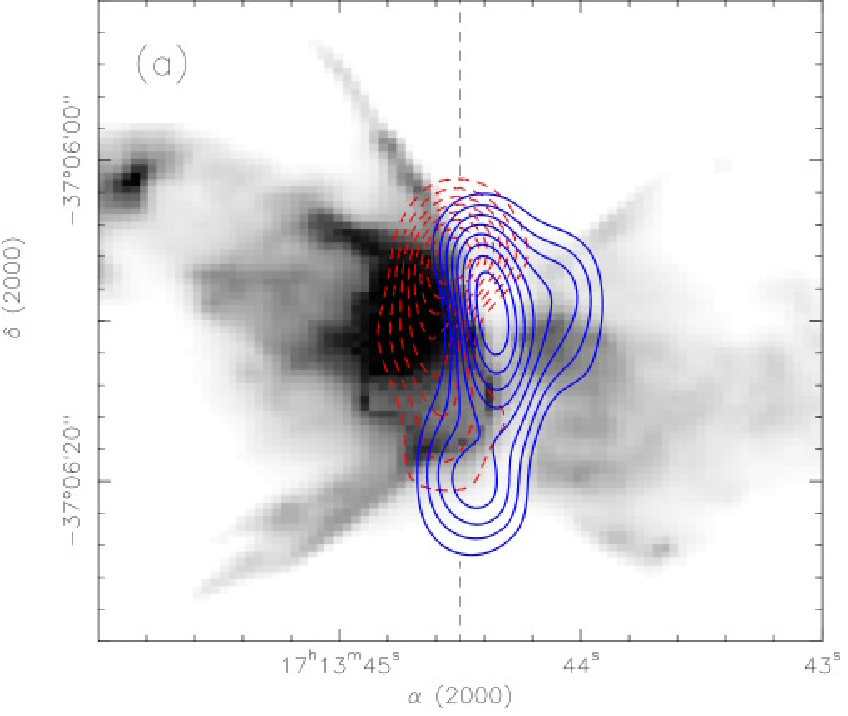}

\hspace{0.7cm}
\includegraphics[width=6.5cm,angle=270]{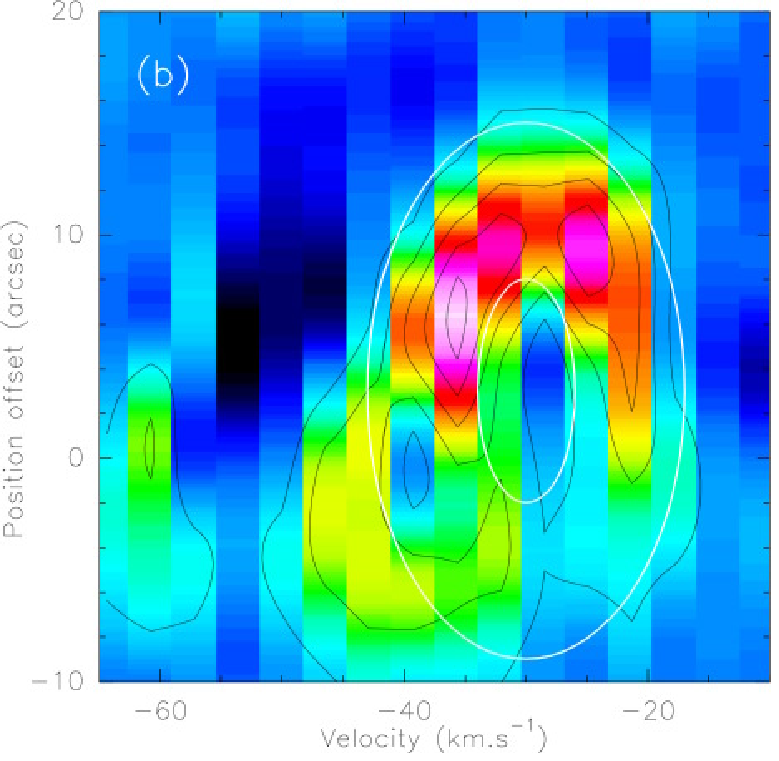}
\caption{\textbf{a)} H$\alpha$ map (grey scale) overlaid with the SMA
  $^{13}$CO \j{2}{1} medium (solid blue lines) and high (dashed red lines)
  components. The black dashed line corresponds to the axis along which we
  constructed the position-velocity diagram shown in (b).
  \textbf{b)} $^{13}$CO \j{2}{1} position-velocity diagram constructed along
  the black dashed axis shown in (a).  The ellipse delimitates
  the molecular emission coming from the torus. The 0\arcsec~position corresponds to $\delta$(2000)=-37$^d$06\arcmin15.9\arcsec.
  \label{ellipse}}
\end{figure}

For the $^{13}$CO \j{2}{1} transition, Fig.~\ref{ellipse}b displays the position-velocity (PV)
diagram constructed along a North-South axis (dashed lines) passing
through the center of NGC~6302. 
In this PV diagram we see the presence of a ring-shape structure
centered at the systemic velocity, $V_{\rm lsr} = -30$~km\,s$^{-1}$, and at
the position $\delta$(2000)=-37$^d$06\arcmin12.5\arcsec. A similar PV diagram is observed in $^{12}$ CO. The low
velocity component on the left part of the $^{13}$CO \j{2}{1} PV diagrams as a fainter peak. The ring-shape
structure observed in the PV diagrams does not suggest rotation as it
would be the case for a rotating disk. Compared to the systemic
velocity the high velocity component, the rear part of the structure,
is red shifted, while the medium velocity component is blue shifted.
This configuration clearly suggests expanding motion. Rather than a
disk, the molecular structure seen around the central star of NGC~6302
is more like an expanding molecular torus.

For a complete, uniform expanding torus seen edge-on, the PV diagram
constructed along the torus should ideally reveal an ellipse whose
extent along the velocity axis directly gives its expansion velocity.
Actually, the shape and overlap of the front and back parts of the
torus (see Fig.~\ref{ellipse}a), as well as the
previous measurements of the NGC~6302 outflow angle compared to the
plane of the sky (i.e. 18\degr; Meaburn et al. 2005) show that this
torus is seen nearly, but not exactly, edge-on. In Fig.~\ref{ellipse}b we plot an ellipse which delimitates the molecular
emission. This ellipse fits quite well the ring-like shape structure,
although we note that there is emission missing in the southern part
of the red shifted emission (i.e. the high velocity component is less
extended than the medium one; cf Fig.~\ref{ellipse}a).  From these
ellipses we can infer the mean projected expansion velocity of
9~km\,s$^{-1}$.  Since we know the angle of the outflow axis to the
plane of the sky, 18\degr, we can, assuming that the torus axis has
the same angle, correct the expansion velocity from the
projection. But since this correction corresponds to only
0.5~km\,s$^{-1}$ which is comparable to the uncertainties, we adopt
V$_{\rm exp} = 9$~km\,s$^{-1}$. If there was some rotation (or overall
velocity gradient) associated with the torus the ellipse in the PV
diagram would be inclined and possibly twisted.  Although difficult to
accurately constrain without a detailed model, Fig.~ \ref{ellipse}b indicates that any such gradient has a magnitude of
$< 20$~km\,s$^{-1}$\,pc$^{-1}$.

The analysis of the lowest velocity CO component is more problematic.
The emission from this component is less similar in the $^{12}$CO
\j{2}{1} and $^{13}$CO \j{2}{1} transitions than is the case for the
torus emission.  However its greater velocity from the systemic
velocity of the sources leads us to interpret this component as
associated with the outflow from the source. Interestingly this
component is projected onto the source in the region where the torus
emission is weakest, perhaps suggesting this may material being
accelerated out of the torus.

\subsection{Position-offset diagrams: ballistic flow}

To uncover the general velocity pattern, we determine for each
velocity channel the position and flux of the brightest pixel.  The
radial offset of this pixel from the stellar position is measured. We
assume the star to be located at the position proposed by Matsuura et
al. (2005a; cf Fig.\ref{ratio_map}).  This technique reduces
drastically the amount of information contained in the images, but has
been found to provide a powerful tool, used, e.g., in maser studies of
circumstellar envelopes.

\begin{figure*}[!t!]
\hspace{1cm}
\includegraphics[width=16cm,clip]{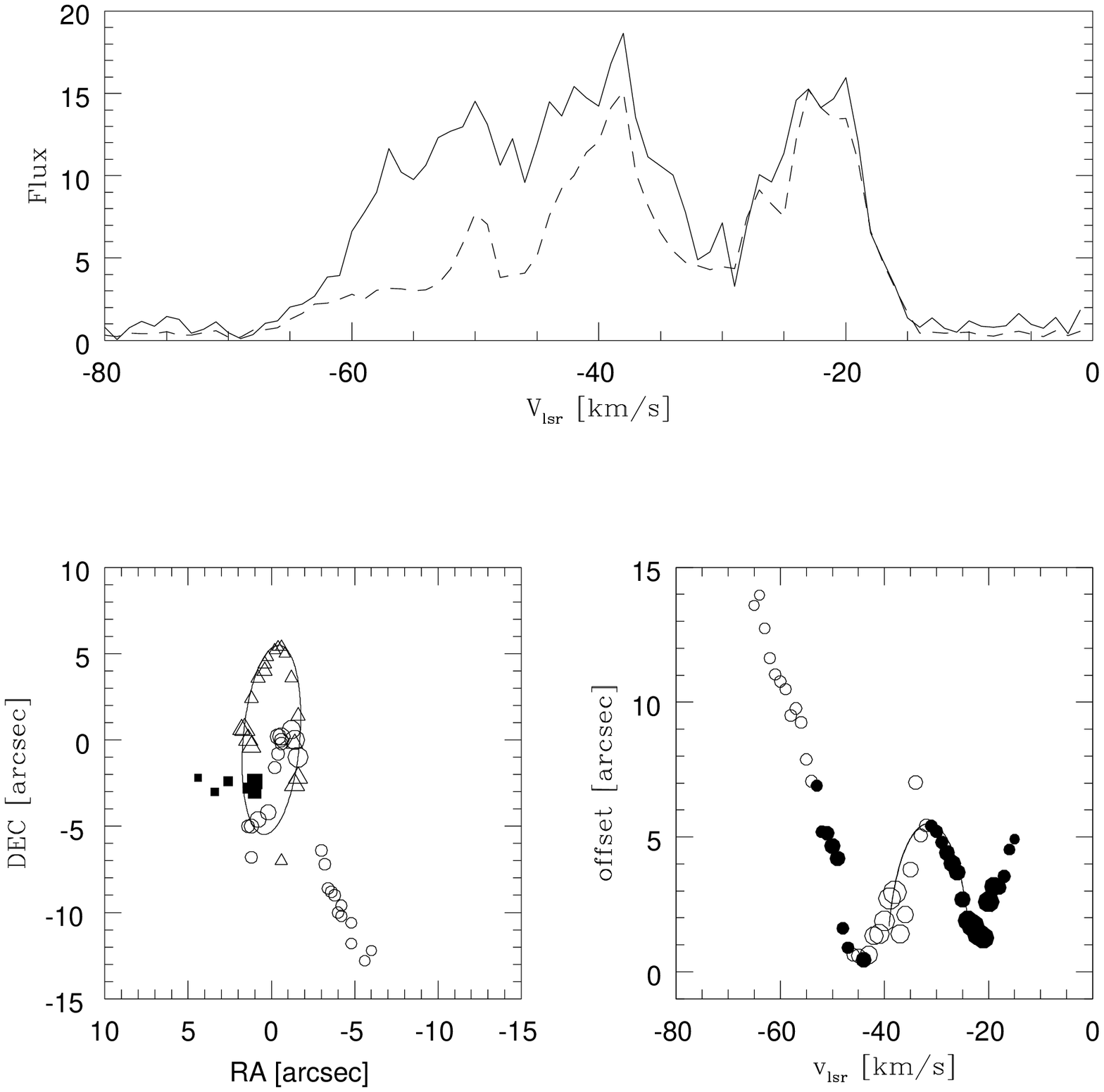}
\caption{Position-offset diagram. Left panel:
positions of the peaks.  Triangles: $ -20 >V> -40\,\rm km\,s^{-1}$;
circles: $V\le-40$; (filled) squares: $V\ge-20\,\rm km\,s^{-1}$.  The
size of the symbols are a (weak) indicator of peak flux.  Only peak
flux values $>1$ mJy/beam (on scale of top panel) are included. The
solid line indicates the model for a tilted ring, with a radius of
5.5\arcsec\, an inclination angle of 18\degr\ compared to the plane of
the sky, and rotated to a position angle of -5\degr. The (0\arcsec,0\arcsec)
position corresponds to the location where the central star is assumed
to be.  Right panel: the velocity versus the radial offset to
the central star.  Filled symbols are positions to the East of the
star.  The solid line is still our best fit torus model where $V_{\rm
lsr}=-31.5\,\rm km\,s^{-1}$ and $V_{\rm exp}=8\,\rm km\,s^{-1}$.
\label{velpos}}
\end{figure*}

We apply this method to the SMA $^{12}$CO \j{2}{1} data, which shows best
signal to noise ratio.  The result is shown in Fig. \ref{velpos}. 
We select all channels with peak flux above 1 mJy/beam. The left panel
shows the positions of these peaks. The stellar position is taken as
(0,0). Different symbols indicate the velocity ranges. Larger symbols
are used for stronger peak fluxes.  The distribution is reasonably
well fitted with a tilted ring, with a radius of 5.5\arcsec, an
inclination of 18\degr\ (Meaburn et al. 2005), rotated to a position
angle of $-5$\degr.

The right panel of Fig. \ref{velpos} shows the velocity versus the
angular separation between the peak and the assumed stellar position.
Such diagrams can distinguish between expanding shells and polar flows
(Zijlstra et al. 2001). Both are present here. The solid line is our
model for the ring, where the best fit is found for a central velocity
$V_{\rm lsr}=-31.5\,\rm km\,s^{-1}$ and an expansion velocity $V_{\rm
exp}=8\,\rm km\,s^{-1}$. These differ slightly from the earlier fits
but are within the uncertainties.  The symbols clustering around the
drawn line shows that part of the CO emission is well derscribed by an
expanding ring. In the following sections we consider the values
inferred from the best fit model as the reference ones.

On Fig.~\ref{velpos} (right) we also notice that at higher and lower
velocity two linear features appear.  These show velocity linearly
increasing with distance from the star. This is indicative of polar
flows. The ballistic nature of these can be understood as due to the
interaction of a slow and a fast wind, where the swept-up interface
moves at a constant but direction-dependent velocity. This gives rise
to a ballistic flow pattern (Zijlstra et al. 2001), seen in bipolar
OH/IR stars.

The velocity range of this component is small in NGC 6302, compared to
typical for bipolar OH/IR stars. The gradient is $\sim 140 \,\rm
km\,s^{-1}\,pc^{-1}$. But it shows a considerable spatial extent, 2-3
times the size of the ring. The linear flows are seen on either side
of the star, and both approaching and receding components with respect
to the systemic velocity, are present.

\section{Discussion}

\subsection{Comparison with other objects}

Molecular gas is relatively common among planetary nebulae.  It is
seen among the youngest objects where the high density material traps
the dissociation and ionization fronts within the nebulae. As the
nebula expands and dilutes, the molecular regions will be overrun
(Huggins et al. 1996, Woods et al. 2005). In the most recent and the
most sensitive survey, Huggins et al.  (2005) detect CO emission for
40 nebula out of a sample of 110, although only a few are as bright as
NGC\,6302.

 CO torii are known for a number of objects in a similar
  evolutionary phase.  M\,1-16 (Huggins et al. 2000) has an extensive
  envelope of 0.12\,M$_\odot$, around a very compact ionized region.
  NGC\,7027 also has a high-mass ($\sim 1\,\rm M_\odot$) CO shell, but
  its ionized nebula is elliptical and lacks the butterfly shape. A
  closer optical morphological analogy is NGC~2346, which is also a
  butterfly nebula: it shows CO distributed in a clumped expanding
  ring with an estimated mass of 0.1\,M$_\odot$ (Bachiller et al.
  1989).

During the preceding phase of the post-AGB nebulae, where the star has
ceased its mass loss and left the AGB, but the star is still too cool
to cause significant ionization, CO is always present. Such objects
tend to show strong bipolarity; CO velocities up to 200 km\,s$^{-1}$
are seen (Bujarrabal et al. 2001) although the spectra are dominated
by more sedate material. In some cases a Keplerian disk is present
(e.g. Bujarrabal et al. 2005). Woods et al. (2005) show that two
distinct classes of objects exist in this phase, one rich in different
molecules and the other only showing a small subset, but both types
show CO emission. 
The molecule rich class (objects such as CRL\,618, OH\,231.8+4.2)
shows the highest density torii, similar to that in NGC\,6302.

However, NGC\,6302 differs from the other objects in the class, in
that it is more evolved, with a much hotter central star and a high
mass for the molecular torus. The full line width of 50\,km\,s$^{-1}$
is typical for planetary nebulae, althought about half the line width
is due to the bipolar, ballistic flow first detected in the SMA data.

\subsection{Torus properties}

The CO emission in NGC\,6302 comes predominantly from a compact,
expanding torus centred on the exciting star. This CO emission
requires that there are at least two different temperature components
of the CO. These components are similar in temperature to those
identified in the dust emission from this source by Kemper et
al. (2002) suggesting that both dust components are associated with
the torus rather than the outflow lobes as Kemper et al.  suggested
for the warmer dust.  The properties of the CO emission have a number
of important implications for the origin and evolution of this system.

The best estimate of the mass of the torus is $\sim2$~\Msun\ assuming
a $^{13}$CO to $^{12}$CO abundance ratio of 15. This mass is similar
to that inferred from the submillimetre continuum observations of
Matsuura et al. (2005a), but significantly higher than those estimated
from previous observations of CO, e.g. Huggins \& Healy (1989). This
discrepancy probably results from a combination of previous
underestimates of both the integrated intensity of the CO \j{2}{1}\
line and opacity of the transition. In addition to the molecular
material traced by the CO, previous observations have identified
0.05\Msun\ of ionised HI (Gomez et al. 1989), and 0.25\Msun\ in low
excitation atomic gas traced by CII (Castro-Carrizo et al. 2001) (all
scaled to the assumed 1\,kpc distance) associated with the source.  The
CO torus therefore completely dominates the mass of the circumstellar
material associated with NGC\,6302.

The high mass of the CO torus means that NGC\,6302 has the highest
mass of inner circumstellar material of any PN or proto-planetary
nebula (PPN) known.  The survey of PPNe, likely models for the
progenitor to NGC\,6302, by Bujarrabal et al. (2001) found only two
objects with masses of circumstellar comparable or larger than found
in NGC\,6302, but both these objects, AFGL2343 and IRAS+10420, are
classified as yellow hypergiants with broad CO lines without a clear
line core-wing separation, unlike the PPN sources.  However it is has
been suggested that that the distance to these sources have been
overestimated, consequently with much lower masses of circumstellar
material (Josselin \& L\`ebre 2001).

As mentioned above, the lowest velocity CO component is not part of
the expanding torus. Spatially it appears close to the region where
the torus emission is incomplete, suggestive of it being part of the
torus which has been distrupted and accelerated by the outflow from
the central star.  Evidence that this may be occurring comes from the
high excitation lines detected by Feibelman (2001). These lines peak
at velocities between $-54$ km/s and $-64$ \kms\ which would associate
them with this most blueshifted CO material. Meaburn et al. (2005)
also finds significantly blueshifted forbidden N{\sc ii} emission at
close to the position of the lowest velocity CO component.  This clump
of CO material therefore appears to the source of at least some of the
material seen in the ionised outflow.  The velocity of these UV and
optical lines is quite different from the $-23$~\kms\ to $-34$~\kms\
velocities of the IR recombination lines and low excitation atomic
lines (Casassus et al. 2000; Castro-Carrizo et al. 2001), which in
light of the CO torus would appear kinematically to originate from
within the CO torus.

\subsection{Age}
The size and velocity of the CO torus can be used to infer a number of
timescales associated with the system.  The radius of the outer edge
of the torus is 13\arcsec, corresponding to $1.3\times10^4$~AU.  With
the best fit model expansion velocity of $8$~\kms\ this implies that
the event which initiated the expansion of the torus occurred
$\sim$7500 years ago.  From the apparent thickness of the torus,
i.e. 8000 AU, the event which produced the torus lasted for about
$\sim$4600 years; the event ended $\sim2900$ years ago.  These
timescales assume that the limits of the current CO emission trace the
full extent of the original material of the torus and that the
apparent inner edge marks the end of the ejection event.

The ballistic flow cannot be dated as its inclination angle with respect
to the plane of the sky is not known. If we assume it dates from
the time after the torus ejection, we find from the velocity
gradient (140\,km\,s$^{-1}$\,pc$^{-1}$) that $i>30$ degrees, where
$i$ is the angle with the line of sight.

Meaburn et al. (2005) find a younger age from the expansion profile
of the north-western optical lobe, 1.7 arcmin from the star. They date
the lobe to an age of 1900\,yr. Within the uncertainties, this may
be similar to the end of the torus ejection (as seen in NGC\,6537, a similar
nebula to NGC\,6302: Matsuura et al. 2005b). However, the bipolar lobes
may also have formed later.

\subsection{Progenitor star}

Combining the total mass of the circumstellar material currently
associated with the nebula, $\sim2$~\Msun\ , with the estimated current
mass of the central star, 0.7-0.8\Msun\ (Casassus et al. 2000) gives
an absolute lower limit on the mass of the progenitor of the system of
$\sim3$\Msun.  The upper mass limit is 8\,\Msun, based on the fact
that stars more massive than this do not become post-AGB stars, but
instead ignite their carbon core. This mass limit is consistent with
the $\sim4$\,\Msun\ progenitor required to produce the high temperature,
high-mass stellar remnant. Intriguingly, this means that the bulk of
the inital mass of the system is still present close to the star.

Bipolar planetary nebulae have on average higher-mass progenitors,
based on their smaller Galactic scale height of bipolar nebulae
(Corradi \& Schwarz 1995; Stanghellini et al. 2002). This is
consistent with the high derived progenitor mass.  The formation of
bipolar nebulae has been linked to binary progenitors (Soker 1998a)
and the angular momentum in such a system could easily drive the mass
loss needed to generate the torus.  Based on IUE observations,
Feibelman (2001) has claimed that NGC~6302 is a binary system with a
G{\sc V} secondary, however the extensive circumstellar material and
the extremely high extinction towards the central star makes this
questionable.  The UV flux more likely represents scattered light from
the central star (see the discussion in Meaburn et al. 2005).

An infrared excess seen at the position of the star (Matsuura et al. 2005a)
possibly represents a very compact circumstellar nebula.  The excess can be
free-free emission (as in Be stars) or very hot dust. The presence of
circumstellar material located within the central cavity can be related
to binarity (as in post-AGB stars with disks: van Winckel 2002). However,
this is indicative only, and the binary nature of the central star remains
suspected but unproven.

Casassus et al. (2000) noted that NGC\,6302 does not show any evidence
for a hot wind or wind blown cavity.This is surprising given the
evidence for the ballistic flow, and the much faster flows in the
ionized gas (Meaburn et al. 2005). The kinematics of the nebula makes
it likely that a fast stellar wind did exist in the past.

\subsection{Mass Loss and Evolution}

For the progenitor to eject 2\,\Msun\ of material over period of 4600
years (based on the thickness of the torus) implies an average
mass-loss rate of $5\times10^{-4}$\msunyr, concentrated towards the
equatorial plane.  Even adopting the outer radius only of the torus to
estimate the timescale and reducing the mass to minimum measured,
1.4\Msun\, only reduces the mass-loss rate to
$2\times10^{-4}$\,\msunyr.  This mass loss rate is consistent with
those measured towards PPNe, although a factor of a few larger than
typical (Bujarrabal et al. 2001).  Interestingly the mass-loss rate
for NGC\,6302 is very similar to that infered from the similarly
massive, but more spatially extended, CO shell around the hot PN
NGC\,7027 (Masson et al. 1985), although its shell is much less
equatorially condensed.

Models show that mass-loss rates can only approach these values during
the late AGB stage of evolution. However values as high as that seen
for NGC\,6302 are difficult to produce in the models, even for short
periods. The NGC\,6302 mass-loss rate exceeds any value obtained  in
models by Vassiliadis \& Wood (1993) and is barely reached at the very
tip of the distribution of the models by Bloecker (1995).

Mass loss on the AGB is assumed to be driven by radiation pressure on
the dust. The achievable rates are limited by the available momentum
in the stellar radiation field: this limit can be exceeded by a factor
of two if multiple scattering of photons is taking into account
(Vassiliadis \&\ Wood 1993). Using the lower limit to the mass-loss
rate, the expansion velocity derived above and a stellar luminosity
$L=10^4\,\rm L_\odot$, we find that the ratio between the momentum in
the torus and the one available in the stellar radiation, is in
NGC~6302 :

$$
 \beta = \frac{\dot M V_{\rm exp}c}{L} = 7.8
$$

Considering that the mass loss is concentrated towards the equatorial
plane, while the stellar radiation is isotropic, worsens the
discrepancy.  We conclude that, despite the low $V_{\rm exp}$, the
mass-loss history of the torus is difficult to achieve by standard
radiation-pressure-driven mass loss.  A similar argument, but with
much larger discrepancies, is presented by Bujarrabal et
al. (2001). On the other hand, the low expansion velocity is as
expected from a dust--driven wind at very high $\rm \dot M$ (Habing et
al. 1994).

In contrast, the energy requirements are relatively easily satisfied
using a hypothetical binary companion. If we assume a 1\,M$_\odot$
companion in a 1\,AU orbit, and that the envelope of a 3\,M$_\odot$
primary star is at the same radius at the onset of mass loss, a
reduction of the secondary orbital radius to $\sim 0.1$\,AU, whilst
the mass of the primary reduces from 3 to 1\,M$_\odot$, will suffice
to provide the energy to eject the torus. In the final configuration,
the orbital period of the hypothetical binary will be $P\sim
15$\,days. We conclude that binary interaction can be a possible
explanation for the ejection of the torus. A prediction of this would
be the existence of a companion with an orbital period $P\lsim
1\,$month.

The only butterfly-type nebula with a known binary nucleus is the
 above-mentioned NGC\,2346, with a period of 16 days. This object also
 has a CO ring, albeit with considerable lower mass than seen in
 NGC\,6302.  It  has an unusal period: almost all other known
 planetary nebulae with close binary central stars show much shorter
 periods, requiring common envelope evolution (Zijlstra 2006).  A
 period of 16\,days requires binary interaction (as the final orbit is
 smaller than the AGB star) whilst avoiding common envelope evolution
 which would lead to shorter orbits. However, whether this is common
 for the butterfly nebulae is not known.

\section{Summary}

High angular resolution observations of $^{12}$CO \j{2}{1} and
$^{13}$CO \j{2}{1} have detected a massive torus centred on the
exciting star.  The observations indicate a lower limit of 15 for the
abundance ratio $^{12}$CO/$^{13}$CO and so by implication,
$^{12}$C/$^{13}$C.  A detailed analysis of the emission in the
\j{2}{1} and \j{3}{2} transitions of $^{12}$CO and $^{13}$CO indicates
that the emission arises from material at two different
temperatures. In total the torus contains $\sim2$~M$_\odot \pm
1$~M$_{\odot}$ of material, a very similar mass to that previously
derived from observations of the dust continuum emission from the
source. From our best fit model, this torus is expanding with a
velocity of 8~\kms\ which leads to a dynamical age of the torus of
7500 yr, with the inner edge of the torus being ejected $\sim2900$
years ago. This ejection may have coincided with the eruptive event
proposed to have formed the systems bipolar lobes.  We speculate that
the ejection of the inner edge of the torus also terminated the fast
CO outflow which, while typical in the PPN progenitors of PN, is not
detected in NGC~6302. The current mass within the PN implies that the
progenitor of the system had a mass of $>3$M$_\odot$ and during the
formation of the torus sustained a mass loss rate of
$5\times10^{-4}\,\rm M_\odot\,yr^{-1}$

The derived mass-loss history is difficult to fit using mass loss
driven by radiation pressure. We find that binary interaction is a 
plausible mechanism leading to the ejection of the torus.

Ultimately the torus is likely to become photoionised by the central
star, as is happening to the CO shell around NGC\,7027. The photon
dominated region where the UV from the star penetrates the CO torus,
which is traced by the warm CO component observed here and the low
excitation atomic lines observed by Castro-Carrizo et al. (2001), is
likely to support a rich and complex chemistry, making NGC\,6302 an
interesting target for further molecular line observations.

%                                     Two column figure (place early!)
%______________________________________________ Gamma_1 (lg rho, lg e)

%\begin{acknowledgements}
 %     Part of this work was supported by the German
  %    \emph{Deut\-sche For\-schungs\-ge\-mein\-schaft, DFG\/} project
 %     number Ts~17/2--1.
%\end{acknowledgements}

%\begin{thebibliography}{}

%\end{thebibliography}

\end{document}